\documentclass[fleqn,usenatbib]{mnras}

\DeclareRobustCommand{\VAN}[3]{#2}
\let\VANthebibliography\thebibliography
\def\thebibliography{\DeclareRobustCommand{\VAN}[3]{##3}\VANthebibliography}

\usepackage{graphicx}	
\usepackage{amsmath}	
\usepackage{mwe}
\usepackage{caption}
\usepackage{subcaption}

\usepackage{float}
\usepackage{enumitem}
\usepackage[none]{hyphenat}
\usepackage{adjustbox}

\usepackage{newtxtext,newtxmath}

\title[2.5-MHD models of circumstellar discs]{2.5-MHD models of circumstellar discs around FS CMa post-mergers : II. Stationary accretion stage}

\author[Moranchel-Basurto et al.]{
A. Moranchel-Basurto,$^{1}$\thanks{E-mail: amoranchel087@gmail.com}
D. Korčáková,$^{1}$ and R. O. Chametla$^{1}$ \\
$^{1}$Charles University, Faculty of Mathematics and Physics, Astronomical Institute, V Hole$\check{s}$ovi$\check{c}$k\'ach 747/2, 180 00 Prague 8, Czech Republic\\
 }

\date{Accepted XXX. Received YYY; in original form ZZZ}

\pubyear{2023}

\begin{document}
\label{firstpage}
\pagerange{\pageref{firstpage}--\pageref{lastpage}}
\maketitle

\begin{abstract}
We study the star-disc interaction in the presence of the strong magnetic field ($B_\star=6.2kG$) of a slowly rotating star. This situation describes
a post-merger of the spectral type B and has not been previously investigated. We perform a set of resistive and viscosity $2.5D$-magnetohydrodynamical simulations using the PLUTO code. Based on our previous work, we consider the initial gas disc density $\rho_{d0}=10^{-13}\mathrm{gcm}^{-3}$ since it
describes the conditions around IRAS 17449+2320 well. We find that the fall of gas towards the star occurs in the mid-plane, and remarkably, intermittent backflow takes place in the mid-plane in all of our models for $R\geq10R_\star$. However, we do not rule out that the funnel effect may occur and cause the accretion closer to the poles. Also, when larger values of viscosity ($\alpha_\nu=1$) and stellar rotation rate ($\delta_\star=0.2$) are considered, we find that the disc exhibits a thickening which is characteristic of FS~CMa-type stellar objects. Additionally, we find that the poloidal magnetic field lines twist over short periods of time, leading to magnetic reconnection causing coronal heating that could explain the presence of the Raman lines found observationally in several  FS~CMa stars. Lastly, we find the formation of several knots in the magnetic field lines near and in the mid-plane of the disc which produce perturbations in the density and velocity components, as well as the formation of shallow gaps whose position depends on the inflation of the magnetic field lines.  
\end{abstract}




\section{Introduction}

In the recent two decades the theoretical and numerical effort to understand the dynamics of gas during the interaction between the magnetosphere and the inner region of an accretion disc of a rotating-magnetized star has produced important advances for some types of stars such as: classical T~Tauri stars, white dwarfs, neutron
stars, X-ray pulsars, among others \citep{Koldoba_etal2002,Romanova_etal2002,Long_etal2005,Bessolaz07,Kluzniak_Rappaport2007,Zanni2009,ZF2013,Cemelij19,Ireland_etal2021,CemKP2023}. Among the most important results are the bounding of the disc truncation radius value due to the star's magnetosphere and the formation of accretion funnel flows. The latter is implicitly associated with the spin-up torque felt by the star and represents a stationary gas accretion state. However, both outcomes remain in the regime of sun-like mass stars including a weak magnetic field ($B_\star\leq1kG$).

Less studied is the star magnetosphere-disc interaction regime for more massive stars with a strong magnetic field \citep{MKS1997,Moranchel-Basurto23,CemKP2023}. Apart from the expensive computational cost, the study of this type of interaction represents a complex problem itself, as has been well pointed out in \citet[][hereinafter called Paper I]{Moranchel-Basurto23}, since very strong magnetic field coming from the central star can produce a highly dynamic accretion of gas (non-stationary stage). It is worth mentioning that the stationary or non-stationary regime of the accretion depends mostly on the initial conditions and the strength of the magnetic field does not play the crucial role. Therefore, being able to restrict the free parameters to the smallest possible number can make the study of the interaction of the magnetosphere with the inner disc more tractable, at least in numerical models. For instance, \citet{Cemelij19}
studied star-disc magnetospheric interaction (including only a high viscosity in the disc, $\alpha_\nu=1$) by an 'Atlas' of numerical models trying to obtain quasi-stationary solutions which can be applied to accretion discs around of the YSO-type objects with a low magnetic field ($B_\star<1kG$). He found that increasing stellar magnetic field in the simulation, the gas density of the disc increases. Furthermore, he found that in models with a faster rotation of the star, additional solutions are produced: (i) a fast axial outflow, and, (ii) with both the conical and axial outflow, similar to what was reported in \citet{Romanova09}.

With all the above in mind, the parameters obtained from the observations play an important role. Within the regime of massive stars with a strong magnetic field are the objects called FS CMa post-mergers. Particular attention must be paid to the member of this set labeled as IRAS~17449+2320 object. Because in a recent study the magnitude of the magnetic field on the surface of the star was determined \citep[$B_\star=6.2kG$; see][and Paper I for details]{Korcakova22}. So, similar to Paper I, we are interested in studying the star-disc magnetosphere interaction in said regime taking into account the observational parameters for the FS~CMa stars, such as: the mass ($M_\star=6M_\odot$) and stellar radius ($R_\star=3R_\odot$) \citep{Kricek}, and the value of the magnetic field mentioned above. Nevertheless, in this case with the aim of seeing if the stationary accretion stage takes place, unlike Paper I, here we include as free parameters the stellar rotation rate $\delta_\star$, the magnetic diffusivity $\alpha_\mathrm{m}$ and the viscosity $\alpha_\nu$ within the disc in our numerical models.

The paper is laid out as follows. In Section \ref{sec:model} we present the physical model, while the numerical implementations used in our 2.5D magnetohydrodynamical simulations are discussed in section  \ref{sec:numerical}. In Section \ref{sec:results} we present the results of our numerical models. A brief discussion is presented in Section \ref{sec:discussion} and finally concluding remarks can be found in Section \ref{sec:conclusions}.

\section{Physical Model}
\label{sec:model} 
Our physical model is composed of a slightly sub-Keplerian disc around a magnetized rotating star with a dipolar magnetic field configuration. The models presented in this study are numerical solutions of the magnetohydrodynamical (MHD) equations in which the effects of viscosity and resistivity within the disc are taken into account. Apart from the free parameters considered in this study, our physical model follows those presented in Paper I.


\begin{figure}
\begin{subfigure}{0.5\textwidth}
    \includegraphics[width=1.0\textwidth]{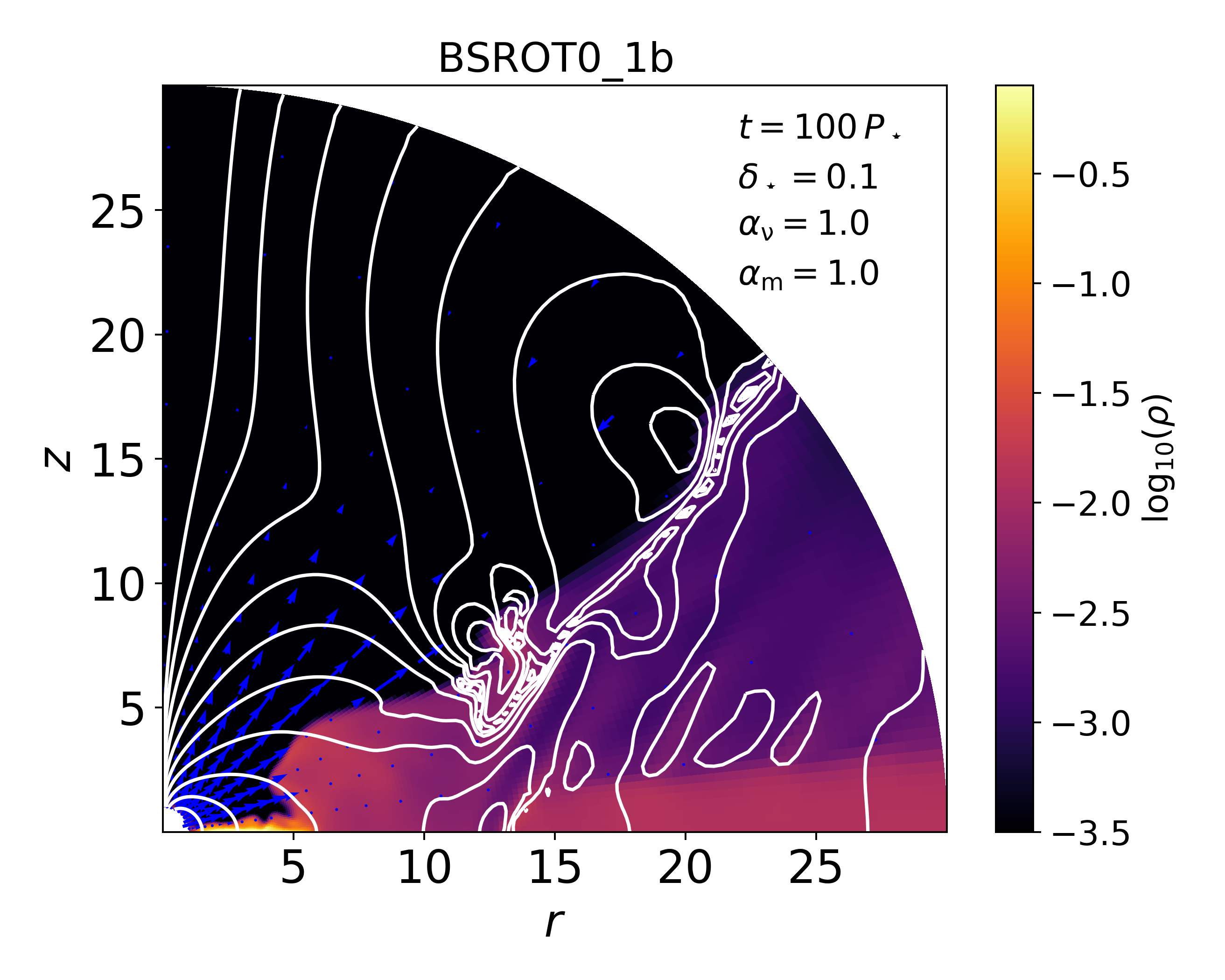}
\end{subfigure}
\begin{subfigure}{0.5\textwidth}
    \includegraphics[width=1.0\textwidth]{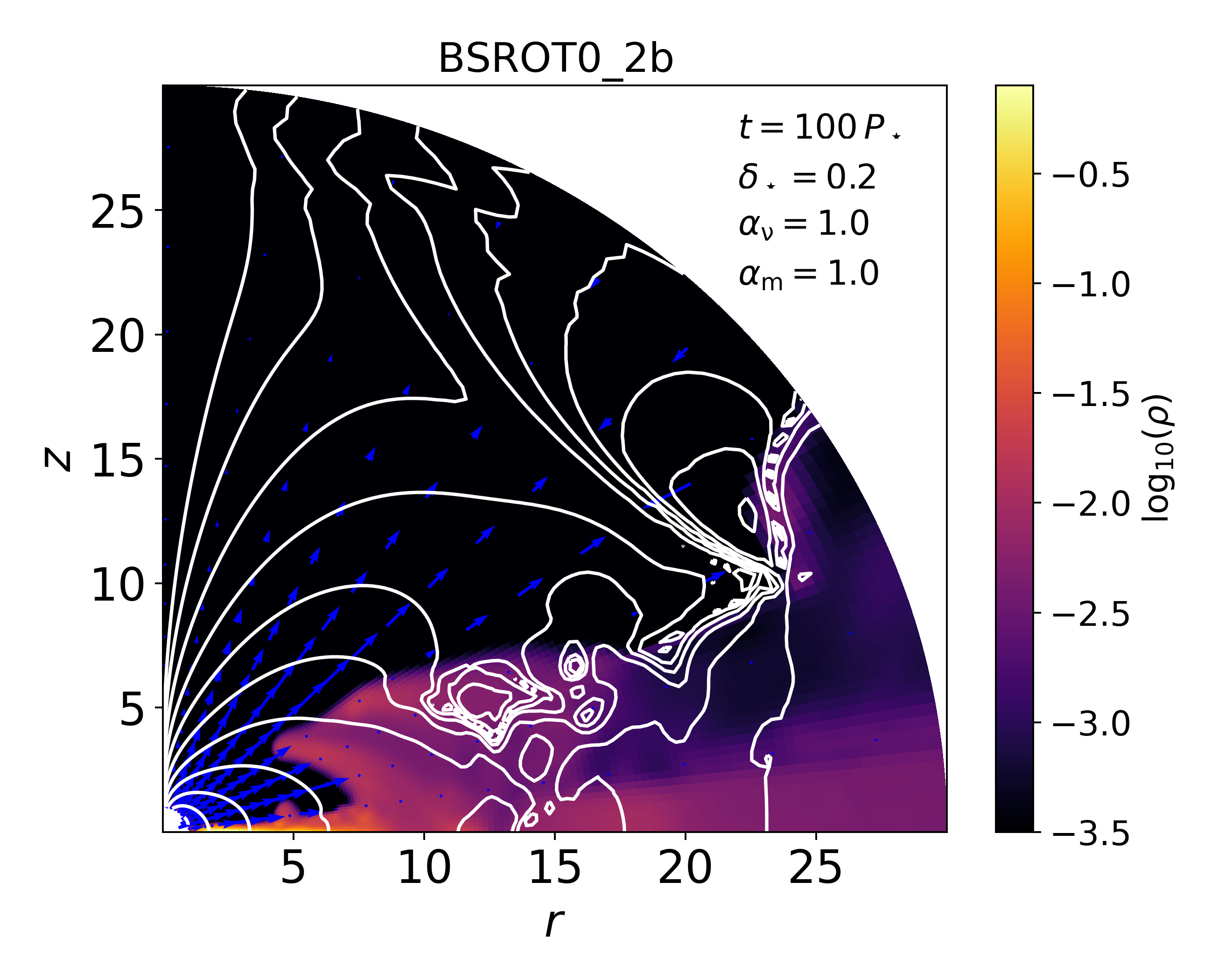}
\end{subfigure}
 
\caption{Upper panel shows logarithmic gas density for the BSROT$\_$1b model with a low rate of stellar rotation $\delta_\star\equiv\Omega_\star/\Omega_\mathrm{br}=0.1$. It can be seen the formation of the funnel effect as well as the formation of disc winds and magnetospheric ejections. Bottom panel shows logarithmic gas density for the BSROT$\_$2b model with a higher rate of stellar rotation $\delta_\star=0.2$. In this case we find the formation of multiple disc layers between $r=5r_0$ and $r=10r_0$ as well as the inflation of the disc that resembles the formation of a huge funnel. The white
lines show the poloidal magnetic field lines. Note that in both models except for the increase in the stellar rotation rate in a factor 2, all other parameters remained fixed.
 }
\label{fig:breaking}
\end{figure}

\subsection{MHD equations}
\label{sec:MHD equations}
The magnetohydrodynamic equations governing the gas dynamics are the conservation of mass
\begin{equation}
    \frac{\partial \rho}{\partial t}+\nabla \cdot (\rho \mathbf{v})=0,
\label{eq:mass}
\end{equation}
the conservation of momentum equation
\begin{equation}
    \frac{\partial \rho \mathbf{v}}{\partial t} + \nabla \cdot \left[\rho \mathbf{v} \mathbf{v} +\left(P+\frac{\textbf{B}\cdot \textbf{B}}{8\pi}\right)\mathbf{I}-\frac{\mathbf{B}\mathbf{B}}{4\pi}- \tau\right]=-\rho \nabla\Phi,
\label{eq:momentum}
\end{equation}
the energy equation
\begin{equation}
\begin{split}
   & \frac{\partial E}{\partial t} + \nabla \cdot \left[ \mathbf{v} \left(E+P+ \frac{\mathbf{B}\cdot\mathbf{B}}{8\pi}\right) - \frac{1}{4\pi}(\mathbf{v}\cdot \mathbf{B})\mathbf{B}\right] \\ & +\nabla \cdot \left(\eta_m \textbf{J}\times \frac{\mathbf{B}}{4\pi} - \mathbf{v}\cdot\tau\right)  = -\rho(\nabla\Phi) \cdot \mathbf{v} - \Lambda_{cool},
  \end{split}   
\label{eq:energy}
\end{equation}
the last term on the left side of the energy equation (\ref{eq:energy}) includes the heating terms, to balance this, note that the cooling term $\Lambda_{cool}$ has been included on the right side of the equation, that means, that the system should evolve adiabatically. Said cooling term also helps prevent the thermal thickening of the gas disc \citep[][]{Zanni2009}.

And finally the induction equation by which the description of the evolution of the magnetic field is described is given by:
\begin{equation}
    \frac{\partial \mathbf{B}} {\partial t}- \nabla\times (\mathbf{v} \times \mathbf{B} - \eta_m \textbf{J})=0.
    \label{eq.induction}
\end{equation}
where $P$ is the thermal gas pressure, $\rho$ the gas density, $\mathbf{v}$
its velocity, $\textbf{I}$ the unit tensor, $\textbf{B}$ the magnetic flux density vector, $\textbf{J}$ is the electric current density given by the Ampére's law $\mathbf{J} = \nabla \times \mathbf{B}/4\pi$, $\eta_m$ is the resistivity and is related with the magnetic diffusivity by $\nu_m=\eta_m/4\pi $, $\Phi = GM_*/R$ the gravitational potential and
$E$ represents the total energy density given by
\begin{equation}
    E=\frac{P}{\gamma-1}+\rho\frac{\mathbf{v}\cdot\mathbf{v}}{2}+\frac{\mathbf{B}\cdot\mathbf{B}}{8\pi},
    \label{eq:dens_energy}
\end{equation}
where $\gamma = 5/3$ is the polytropic index of the plasma. On the other hand,
the viscous stress tensor $\tau$ is defined as:
\begin{equation}
    \tau = \eta_v \left[(\nabla \mathbf{v})+(\nabla\mathbf{v})^T - \frac{2}{3}(\nabla\cdot \mathbf{v})I\right],
\end{equation} 
where $\eta_v $ is the dynamic and $\nu_v= \eta_v/\rho$ the kinematic viscosity, respectively.

\subsection{Accretion disc}
\label{sec:accretion_disc}
The density $\rho_{d}$ and pressure $P_{d}$ of the gas in the accretion disc is set following the three-dimensional models of Keplerian accretion discs considering spherical coordinaties $(R,\theta,\phi)$ used in \citet{Zanni2009}  \citep[which in turn is based on the disc model of][]{KK2000}:
\begin{equation}
    \rho_d (R,r) = \rho_{d0}\left\{\frac{2}{5h^2}\left[\frac{R_0}{R}-\left(1-\frac{5h^2}{2}\right)\frac{R_0}{r}\right]\right\}^{3/2},
    \label{eq:rho_}
\end{equation}

\begin{equation}
    P_d = h^2\rho_{d0}v_{K0}^2\left(\frac{\rho_d}{\rho_{d0}}\right)^{5/3}
\end{equation}
where $h=C_s/v_K$ is the aspect ratio, with $C_s$ and $v_{K}= \sqrt{GM_*/R}$ the sound speed and the Keplerian velocity, respectively. In Eq. (\ref{eq:rho_}) $\rho_{d0}$ and $v_{K0}$ denote the initial values of the density and the Keplerian velocity calculated at $R=R_{0}$ in the mid-plane of the disc. Note that we define $r=R\sin{\theta}$ as the cylindrical radius.

The velocity components $(u_{Rd},u_{\theta d},u_{\phi d})$ of the accretion disc are calculated as \citep{KK2000,Zanni2009,ZF2013,Cemelij19}: 

\begin{equation}
    u_{Rd}=-\alpha_\nu h^2\left[10-\frac{32}{3}\Lambda\alpha_\nu^2-\Lambda\left(5-\frac{1}{h^2\tan^2\theta}\right)\right]\sqrt{\frac{GM_\star}{R\sin^3\theta}},
    \label{eq:Vr}
\end{equation}

\begin{equation}
    u_{\phi d} = \left[\sqrt{1-\frac{5h^2}{2}}+\frac{2}{3}h^2\alpha^2_\nu\Lambda\left(1-\frac{6}{5h^2\tan^2\theta}\right)\right]\sqrt{\frac{GM_*}{r}},
    \label{eq:Vphi}
\end{equation}

\begin{equation}
    u_{\theta d}=0.
\end{equation}
In Eqs (\ref{eq:Vr}-\ref{eq:Vphi}), $\Lambda=(11/5)/(1+64\alpha_\nu^2/25)$. Note that we are considering a slightly sub-Keplerian disc.

\subsection{Disc atmosphere}

In a similar form to \citet{Zanni2009}, we included a non-rotating polytropic  hydrostatic atmosphere with a density 
\begin{equation}
    \rho_{\mathrm{atm}} (R)=\rho_{\mathrm{atm}}^0\left(\frac{R_*}{R}\right)^{\frac{1}{\gamma-1}}
\end{equation}
and pressure
\begin{equation}
    P_{\mathrm{atm}}(R)=\rho_{\mathrm{atm}}^0\frac{\gamma-1}{\gamma}\frac{GM_\star}{R_*}\left(\frac{R_*}{R}\right)^{\frac{\gamma}{\gamma-1}}
\end{equation}
with $\gamma =5/3$.
The density contrast between the disc and the atmosphere is $\rho_\mathrm{atm}^0/\rho_{d0}=0.01$, which is kept fixed in all our models.

\subsection{Magnetic field configuration}
We assume that the star is located at the origin of the coordinate system.
The stellar magnetosphere is modeled initially as a purely dipolar field aligned with the rotation axis of the star-disc system. The magnetic field is defined by the flux function $\Psi_*$
\begin{equation}
    \Psi_*(R,\theta)=B_*R_*^3\frac{\sin^2{\theta}}{R}
\end{equation}
where $B_*$ is the magnetic field at $R_{*}$ and $z=0$. The radial and polar field components are therefore given respectively by

\begin{equation}
    B_R = \frac{1}{R^2\sin{\theta}}\frac{\partial\Psi_*}{\partial\theta}
\end{equation}
and
\begin{equation}
    B_\theta = -\frac{1}{R\sin{\theta}}\frac{\partial\Psi_*}{\partial R}.
\end{equation}
The relation between flux function and potencial vector is given by:
\begin{equation}
    \Psi_*=RA_\phi\sin{\theta}
\end{equation}
and then the components of potential vector in spherical coordinates are given by:
\begin{equation}
    A_\phi(R,\theta)=\frac{B_*R_*^3\sin{\theta}}{R^2},
\end{equation}
 $A_R=0$ and $A_{\theta}=0$.

It should be noted that for a suitable numerical treatment of the magnetic field, we employ the “field-splitting” technique \citep{Tanaka1994,Powell_etal1999} which is widely used in this type of studies \citep[see][and references therein]{Cemelij19}.


\begin{figure}
\begin{subfigure}{0.5\textwidth}
    \includegraphics[width=1.0\textwidth]{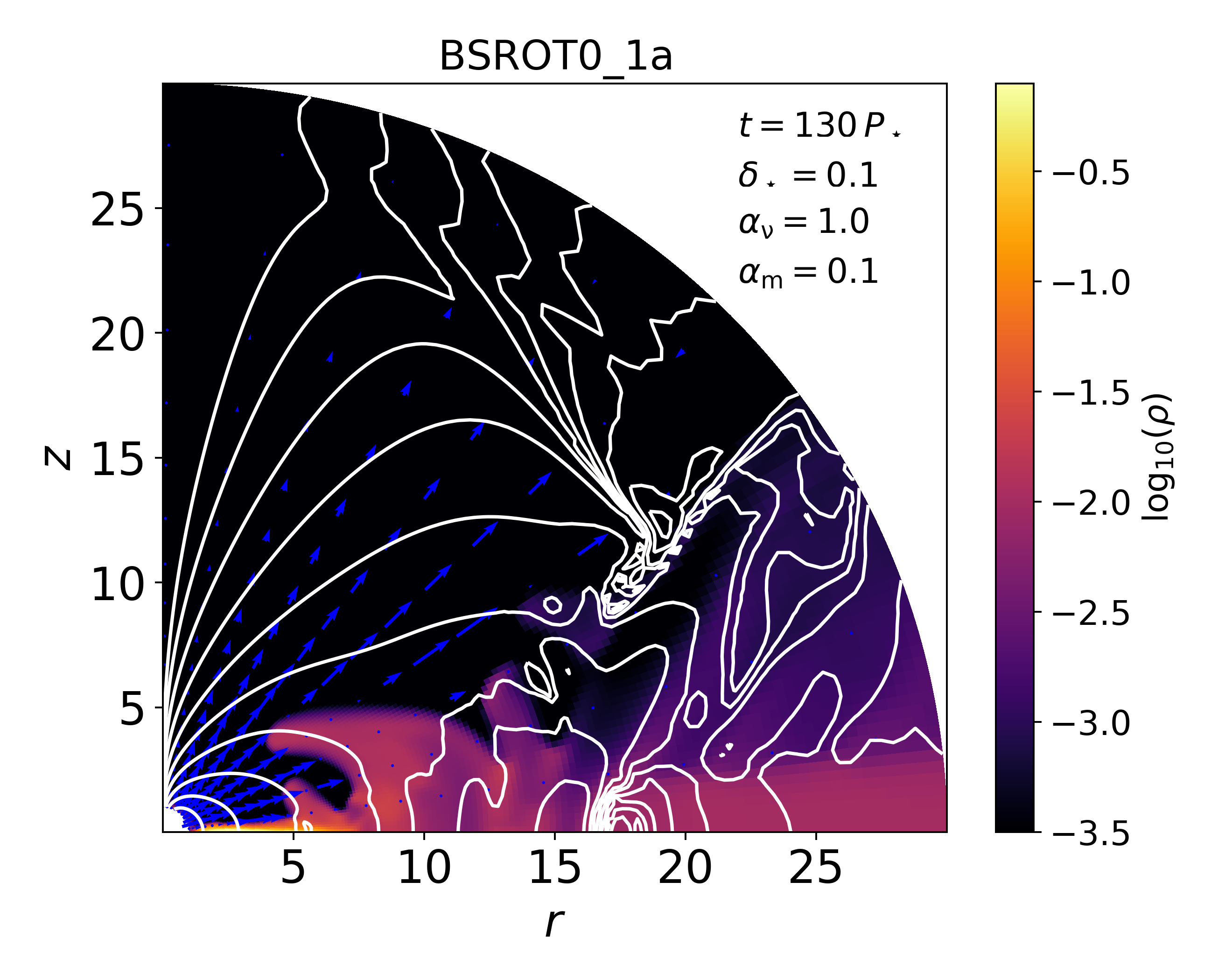}
\end{subfigure}
\begin{subfigure}{0.5\textwidth}
    \includegraphics[width=1.0\textwidth]{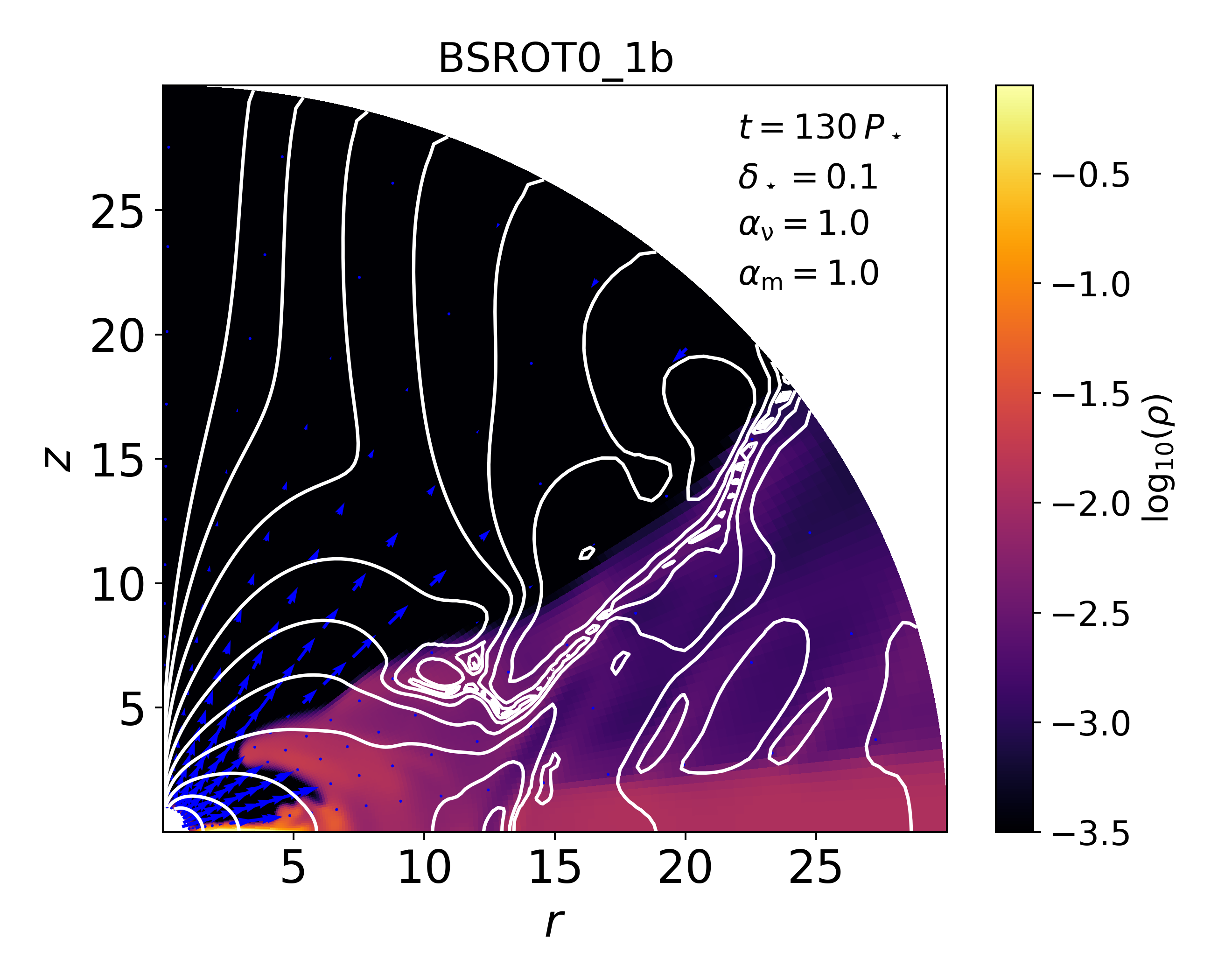}
\end{subfigure}
 
\caption{Upper panel shows logarithmic gas density for the BSROT$\_$1a model with a low rate of stellar rotation $\delta_\star=0.1$ as well as a low magnetic diffusivity $\alpha_m=0.1$. Bottom panel shows logarithmic gas density for the BSROT$\_$1b model with a higher magnetic diffusivity $\alpha_m=1.0$. The white lines show the poloidal magnetic field lines.}
\label{fig:diff}
\end{figure}

\subsection{Magnetic diffusivity models}
\label{sec:2.5}
The role of magnetic difussivity $\nu_m$ can be important, this can arise from turbulence generated in the disc and this parameter may be responsible for modifying the strength of the magnetic field lines. For this reason we decided to explore different descriptions of resistivity $\eta_m =4\pi\nu_m$, considering that the magnetic diffusivity $\nu_{m}$ is given by:

\begin{itemize}
\item Classical description \citep[][]{Zanni2009,Cemelij19}:
\begin{equation}
  \nu_m=\frac{\alpha_m C_s^2}{\Omega_K} 
  \label{eq:nuC}
\end{equation}
\item Modified first version \citep[see][and references therein]{MA2022}
\begin{equation}
    \nu_m=\frac{\alpha_m C_s^2}{\Omega_K} \sqrt{1+\beta_\phi}
    \label{eq:nuM}
\end{equation}

    \item Radial and vertical dependence as given in \citet{Bessolaz07}:
    \begin{equation}
        \nu_{m} = \alpha_{m} \Omega_K H^2 \exp\left[-\left(\frac{z}{H}\right)^4\right] ,
        \label{eq:Bess}
    \end{equation}

\end{itemize}
the latter decreasing on a disc scale height $H=C_s/\Omega_K$, here $\Omega_K$ is the Keplerian rotation speed. In Eq. (\ref{eq:nuM}), $\beta_\phi$ is a constant that we fixed to the value 0.1. Throughout this study we will focus in more detail on the resistivity given in Eq. (\ref{eq:Bess}), since this expression takes into account the vertical stratification of the accretion disc. We have taken the $z$-axis parallel to the rotation axis of the disc and assume that the magnetic dipole moment is aligned with the rotational axis.

\section{Numerical Implementation}
\label{sec:numerical}
\subsection{Code}
\label{subsec:code}

In order to investigate the disc-star interaction, we use the MHD module provided by PLUTO code version 4.4 (\citealt{Mignone2009}, \citealt{Mignone2007}) to perform adiabatic global MHD simulations of accretion discs to numerically solve the MHD equations given in section \ref{sec:MHD equations}. Simulations runs on spherical geometry $(R,\theta,\phi)$, were performed using the second-order linear reconstruction Runge Kutta
integration in time to advance conserved variables. To enhance the stability of the code, in the subroutine pml\_states we consider the Van Leer limiter in the density and in the magnetic field since is more diffusive, while a monotonized central differences limiter (minmod) was used in pressure and velocity fields. Also, we mention that for the inclusion of the cooling term that controls the possible heating of the disc, we have used a power-law cooling option provided in the code. Additionally, inspired by the work of \citet{Cemelij19} we use a hll solver with a modification in the flag\_shock subroutine: flags were set to switch to a more diffusive hll solver when the internal energy defined by $e=P_\mathrm{gas}/(\gamma -1)$ was lower than $1\%$ of the total energy given by the equation (\ref{eq:dens_energy}). Furthermore the condition $\nabla \cdot B=0$ was maintained using the constrained transport method. Lastly, we mention that in this work we have followed closely the initial conditions in the disc and corona and boundary conditions at the edges of the computational domain as in the works of \cite{Cemelij19} and \cite{Zanni2009}, in the next subsections we described these conditions in more detail. 

\begin{figure}
\includegraphics[width=\columnwidth]{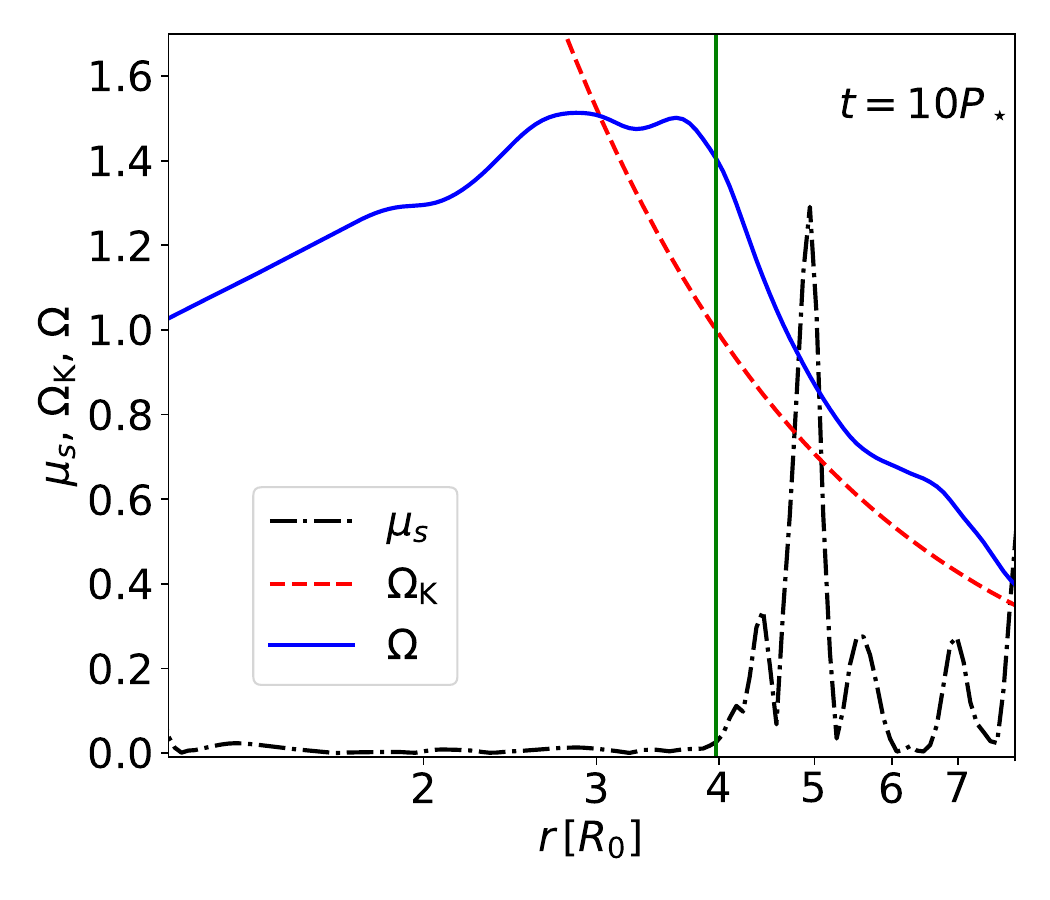}
    \caption{Radial profiles at the midplane of the Sonic Mach number, angular velocity in the disc $\Omega$ (normalized by $\Omega_\star$) and the Keplerian velocity $\Omega_\mathrm{K}$ for the the BSROT$\_$1a model calculated at $t=10P_\star$. The green vertical line shows the position of the truncation radius given in Eq. (\ref{eq:Rt}).}
    \label{fig:Ms}
\end{figure}


\subsection{Simulation parameters and units}
\label{secc:parameter_units}
The parameters of the simulations are the same as were used in the numerical models of the paper I, and were selected to match a set of observed values for the FS~CMa stars
\citep{Korcakova22}. 
More specifically, we fix the mass, radius, initial disc density and the magnetic field of the star to be $M_*=6M_\odot$, 
$R_*=3R_\odot$, $\rho = 1\times10^{-13} \mathrm{g\, cm^{-3}}$ and $B_*=6.2\times10^{3}G$. Furthermore we take into account that the aspect ratio is $h=0.1$. The different models considered are presented in Table \ref{tab:Models}.

\begin{figure}
    
\begin{subfigure}{0.5\textwidth}
    \includegraphics[width=0.90\textwidth]{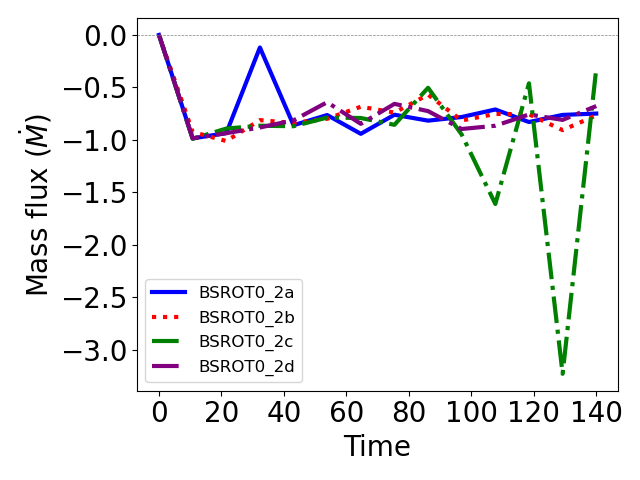}
\end{subfigure}
\begin{subfigure}{0.5\textwidth}
    \includegraphics[width=0.90\textwidth]{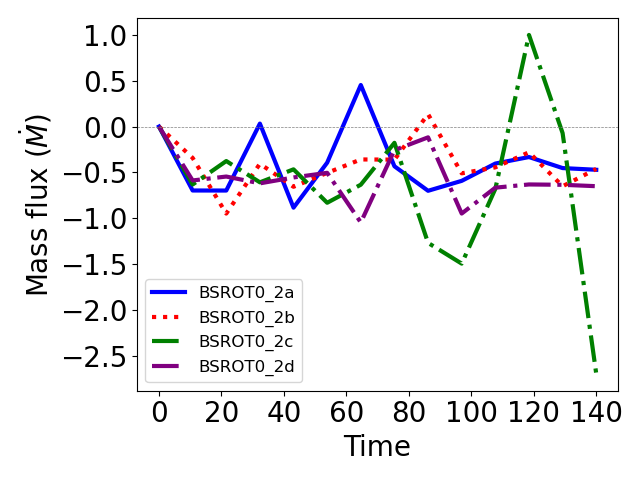}
\end{subfigure}
\begin{subfigure}{0.5\textwidth}
\includegraphics[width=0.90\textwidth]{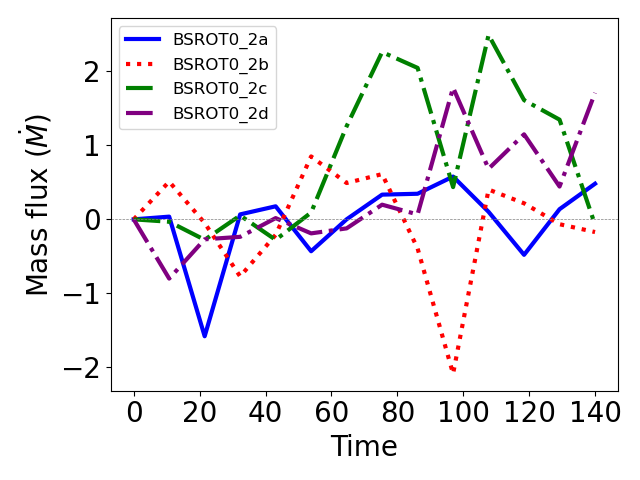}
\end{subfigure}
\caption{Time dependence of the mass flow rate $\dot{M}$, for the BSROT0$\_$2 models, calculated at $R= 3.0R_0$ (\textit{top panel}), $R=3.96R_0$ (\textit{middle panel}) and $R=10R_0$ (\textit{bottom panel}), respectively.}
\label{fig:massflow}
\end{figure}
We use the following dimensionless variables:
\begin{align*}
    R^{\prime}&=R/R_0 & \rho'& = \rho/\rho_{0}  & v'& =v/v_{K0} \\
    t'&=t/(r_0/v_{K0}) & p'&=p/(\rho_{0}v^2_{K0}) &          
B'&=B/(4\pi\rho_{0}v^2_{K0})^{1/2}
\end{align*} 
with
$R_0=R_*=2.087\times 10^{11}\mathrm{cm}$, $v_{K0}=\sqrt{GM_*/R_0}\approx 6.19 \times 10^7 \mathrm{cm\ s^{-1}}$, which is
the Keplerian velocity at $R_0$. For the reference density we take $\rho_{0} = 1\times10^{-13} \mathrm{g\, cm^{-3}}$.
 
In these units, the reference angular rotation rate is $\omega_0=v_0/R_0 =2.96\times 10^{-4}\mathrm{s^{-1}}$, the corresponding timescale is
$t_0= R_0/v_0 \approx 3.37\times 10^3\mathrm{s}$ and the rotation period at $R =R_0$, $T_0=2\pi t_0 \approx 0.245$ days.
For simplicity, we will omit the primes from now on.

Once the initial conditions are normalized, the problem depends on six dimensionless parameters: the aspect radio $h$, the equatorial stellar field intensity $B_*/B_0$, the stellar rotation rate $\delta_*=R_* \Omega_*/V_{K*}$, the coronal density contrast $\rho_\mathrm{atm}/ \rho_{d0}$, and the viscous and resistive coefficients $\alpha_{\nu}$ and $\alpha_m$. Except for the transport coefficients and the initial differential rotation of the star, the other parameters are the same as were used in the simulations of paper I: $h=0.1$, $B_*=6200G$, $\rho_\mathrm{atm}/\rho_{d0}=1\times 10^{-2}$.

\subsubsection{Co-rotation and truncation radius}

To study the interaction of a star-magnetized disc it is important to identify where the truncation radius must occur by the stellar magnetosphere. So far there are numerous studies that debate the most precise way to determine this radius \citep[see for instance:][]{Romanova_etal2002,   Zanni2009, Bessolaz07, Long_etal2005}. However, despite the differences between all the models, it has been determined that for accretion to occur the truncation radius $r_t$ must be located within the co-rotation radius $r_\mathrm{co}=\left(GM_\star/\Omega_\star\right)^{1/3}$. We can rewrite the co-rotation radius in terms of the star radius $R_\star$ if we isolate $\Omega_\star$ variable from the definition of stellar rotation rate $\delta_\star$ given in the previous section and substitute its value in the definition mentioned above, we will obtain after doing a bit of algebra that the radius of co-rotation can be written as follows $r_\mathrm{co}= 4.64R_\star$.

On the other hand, we obtained the value for the truncation radius following the next expression \citep{Bessolaz07}:

\begin{equation}
    \frac{r_t}{R_\star}\approx 2m_s^{2/7}B_\star^{4/7}\dot{M}_a^{-2/7}M_\star^{-1/7}R_\star^{5/7}
\end{equation}
considering that the value of $m_s$ can be close to 1 and substituting the values described in section (\ref{secc:parameter_units}) we obtain that:
\begin{equation}
  \frac{r_t}{R_\star}\approx 3.97121258.   
  \label{eq:Rt}
\end{equation}

It should be emphasized that in the calculation of $r_t$ given in Eq. (\ref{eq:Rt}), we have considered an accretion rate $\dot{M}=10^{-8}M_\odot \mathrm{yr}^{-1}$, which is a conservative value because there is currently debate about what is the correct value of the accretion rate  in the FS CMa-type objects \citep[see][and references therein]{Carciofi_etal2010}.

\subsection{Mesh domain}

The axisymmetric disc is modeled in the radial range $1R_0\leq R\leq 30R_0$ with a polar angular extension in one quadrant ($0\leq\theta\leq\frac{\pi}{2}$). All of our numerical simulations presented here were performed by the same numerical resolution, with a logarithmically increasing grid in radial direction $N_R=109$ and, $N_\theta=50$ polar grid cells uniformly distributed. The resulting computational mesh is sufficient to resolve properly the star-disc magnetospheric interaction (SDMI) in both dimensions \citep[see][]{Cemelij17}, and in turn allows us to study a larger number of parameters (see Table \ref{tab:Models}) with a reasonable computational cost.

\subsubsection{Boundary Conditions and internal boundary}

Since we are interested in analyzing the gas dynamics when SDMI occurs, considering a slow-rotating massive star with strong magnetic field (see subsection \ref{secc:parameter_units}), we take particular care about the boundary conditions implemented on the stellar surface as in previous studies where the SDMI has been analyzed for young stars with a weaker magnetic field \citep[see for instance][]{Romanova_etal2002,Romanova09,Zanni2009,Cemelij19}. Then, the boundary conditions that were used in carrying out this work closely follow those described in the works of \citet{Zanni2009} and \citet{Cemelij19}.

To obtain adequate gas dynamics on the surface of the star, we implement the following conditions. First, in the reference frame co-rotating with the star, the electric field is zero \citep[see equation (11) in][for details]{Zanni2009}. Second, to prevent low-density near the star, we modified the density value when it falls below some limit value\footnote{The low-density limit used here is $\rho=5\times10^{-8}$ in code units, similar to the value used in \citet{Cemelij19}.} within a grid cell of the computational domain on the surface of the star. For the corona region, to conserve the same sound speed and to keep the conservation of the momentum we modify the gas pressure and the velocities, respectively. Lastly, to follow the evolution of the disc, we activate the scalar tracer, taking a value of unity in the disc domain and zero outside of it. Note that because the resistivity described in the equations (\ref{eq:nuC})-(\ref{eq:Bess}) is only applied within the domain of the disc mesh, the reconnection of the magnetic field arising in the magnetosphere is handled by numerical dissipation.

For the inner grid ghost cells, following \citet{Cemelij19} we prescribe the values of the density, pressure, and toroidal components of the velocity and magnetic field from the active zones. This implies that the density and the magnetic field are linearly extrapolated considering a Van Leer limiter, and minmod limiter in the pressure and velocity (see subsection \ref{subsec:code}). It should be emphasized that to achieve numerical stability in the corona, when $v_R > 0$ in the inner radial boundary, we define the gas pressure as $P=(2/5-T_fv^2)\rho_cR^{-5/2}$, with $T_f$ a free parameter \citep[see][]{Cemelij19}. 

On the other hand, in the radial outer boundary condition, we consider the outflow condition for the velocity components, and for the density and magnetic field a linear extrapolation was used. Lastly, in the $\theta$-direction we use the PLUTO code options 'axisymmetric' and 'eqtsymmetric' for the boundary conditions at $\theta=\theta_\mathrm{min}$ and $\theta=\theta_\mathrm{max}$, respectively, since we simulate only one hemisphere of the disc.

\section{Results}
\label{sec:results}

In this section we present the results of our $2.5D$-MHD simulations of a resistive accretion disc for different values of $\alpha_\nu$ and $\alpha_\mathrm{m}$, and also considering different stellar rotation rates. Our study is focused on the strong magnetic field regime measured in massive stellar objects such as FS CMa-type stars \citep[see][]{Korcakova22}.

All the parameters and the different models are given in the table \ref{tab:Models}. The viscosity parameter describes the strenght of the viscous torque, that allows the disc to accrete, while the resistivity parameter $\alpha_m$ allows coupling of the stellar magnetic field with the disc material.

For both viscosity and resistivity we consider the pair of values of 0.1 and 1. So the Prandtl number can take the set of values of $P_\mathrm{m}=\{0.06,0.6,6.6\}$. The motivation for choosing these values in the viscosity lies in the fact that previous studies of the star-disc interaction show that if it is fulfilled $\alpha<\alpha_\mathrm{crit}\simeq0.685$, a backflow is originated in the mid-plane of the disc, which results in the accretion of the disc towards its surface \citep[see][and references therein]{Zanni2009}.

On the other hand, the choice of values 0.1 and 1 for the $\alpha_\mathrm{m}$ resistivity coefficient, are motivated to know which region of the disc is steadily connected to the star in our models with a strong magnetic field. Since the opening of the magnetosphere is intrinsically driven by the differential rotation between the disc and the star, as well as with the resulting buildup of toroidal magnetic pressure. The latter can be modified by the value that $\alpha_\mathrm{m}$ takes on the disc, since if $\alpha_\mathrm{m}=1$ limits the growth of a toroidal magnetic field, producing a larger connected region (an extended magnetosphere). In the case when $\alpha_\mathrm{m}=0.1$ it has the opposite effect on the magnetic pressure (due to the strong coupling), thus producing a smaller connected region (compact magnetosphere).

\subsection{Comparison of the stellar rotation rate for models with an extended magnetosphere}

Figure \ref{fig:breaking} shows a comparison between the BSROT0$\_$1b and BSROT0$\_$2b models. In these models the only difference in the parameters is a factor of 2 in the rate of stellar rotation $\delta_\star$. We find that in the case of a slower stellar rotation (BSROT0$\_$1b model) it produces an accretion towards the star in the mid-plane of the disc as well as a distribution of the gas of the disc outside the mid-plane that resembles the formation of the funnel effect that can give rise to accretion near the poles. Also, we find in this case the formation of disc winds and magnetospheric ejections. In the case of T Tauri stars with a smaller magnetic field on the star's surface ($B_\star=1.1kG$) a similar behavior has been found in the formation of the funnel effect and in the generation of magnetosphere ejections \citep[see][]{ZF2013}. However, in said study there are no signs of gas accretion in the mid-plane of the disc. In the case of a star rotating a little faster (case described here by the BSROT$\_$2b model), we find an inflation of the disc, which is the result of the inflation and reconnection of the magnetic field lines, in addition to the fact that we find the formation of layers between $r=5R_0$ and $r=10R_0$. These layers fail to form multiple funnels towards the star due to the increase in the rotation of the star.

\subsection{Models with compact and extended magnetospheres with the same stellar rotation rate}

In the Fig. \ref{fig:diff} we show the logarithmic gas density maps for the BSROT0$\_$1a and BSROT0$\_$1b models calculated at $t=60P_\star$. In the first case, we find that in a disc with lower magnetic diffusivity ($\alpha_\mathrm{m}=0.1$) there is a formation of two knots at $R=17$ and $R=25R_0$. The first knot creates strong disturbances in the density of the disc, producing a partial gap in $R=17R_0$. The second knot located outside the mid-plane of the disc (at $R=25R_0$) generates a magnetosphere ejection. For the BSROT0$\_$1b model with a higher resistivity ($\alpha_\mathrm{m}=1$), we find the formation of a single knot outside the mid-plane of the disc at $R=13R_0$ which produces a magnetosphere ejection. In both models we find that the gas accretion takes place in the mid-plane of the disc. We emphasize that, since the stellar rotation rate in these models is the same, it is clear that the differences in the gas density distribution arise due to the value of the resistivity within the disc. 

Figure \ref{fig:Ms} shows the radial profiles of the sonic Mach number $\mu_s= {|u_r|/\sqrt{P/\rho}}|_{z=0}$, the disc angular velocity $\Omega$ and the Keplerian velocity $\Omega_\mathrm{K}$, respectively (nomalized by $\Omega_\star$), calculated at $t=10P_\star$ for the BSROT0$\_$1a model. The behaivor of the dynamics in the disc is changing the accretion Mach number which increases and reaches a maximum value of $\mu_s\approx1.3$ in a radius of $5R_0$  (see Fig. \ref{fig:Ms}). Note that $\mu_s$ starts to increase outward from the truncation radius $r_t$ (green vertical line in Fig. \ref{fig:Ms}). On the other hand, we can see in Fig. \ref{fig:Ms} that the angular velocity of the disc $\Omega$ becomes super-Keplerian for $R>3.1R_0$ (solid blue line). This implicitly means that the rotating star exerts a positive torque on the disc. In other words, the star loses angular momentum which consequently translates into a lower rotation speed of the star. Although in this case, the magnetic field is strongly coupled to the disc (since $\alpha_\mathrm{m}=0.1$), we point out that we find very similar results in cases when the resistivity is higher ($\alpha_\mathrm{m}=1$). Therefore, if the magnetic field is able to couple to the disc, the rotation speed of the star will inevitably decrease. The latter may explain why FS CMa-type stellar objects as IRAS 17449+2320 with a strong magnetic field exhibit a very low stellar rotation rate.

\subsubsection{Magnetospheric ejections}

In previous studies of disc models that include a compact magnetosphere ($\alpha_m=0.1$) and a weak magnetic field, magnetospheric ejections have been found \citep[][]{ZF2013,Cemelij19,CemKP2023}. This type of coronal ejections can substantially modify the stellar rotation depending on the strength of the magnetic field and can produce a spin-down of the star \citep[][]{CemKP2023}. Remarkably, we find magnetospheric ejections in both disc models with a compact and with an extended magnetosphere (that is, $\alpha_m=0.1$ and $\alpha_m=1.0$, see for instance Fig. \ref{fig:diff}). We think that in the case of FS CMa-type stars the observed low-stellar rotation is produced by such magnetospheric ejections and, therefore, definitely deserves a detailed investigation which is beyond the scope of this study and should be studied elsewhere.

\begin{figure}
\begin{subfigure}{0.5\textwidth}
    \includegraphics[width=0.83\textwidth]{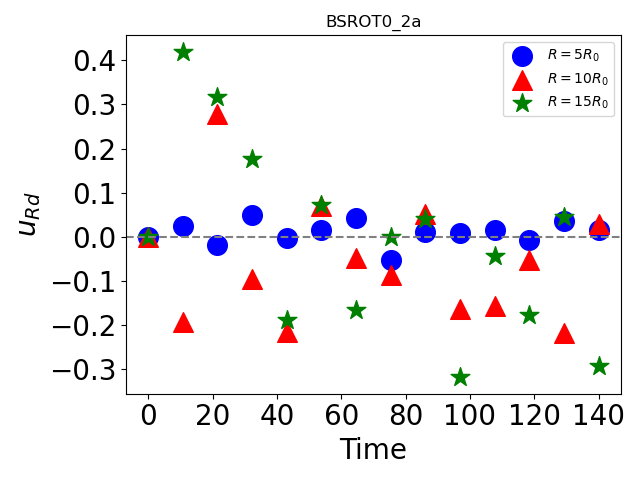}
\end{subfigure}
\begin{subfigure}{0.5\textwidth}
    \includegraphics[width=0.83\textwidth]{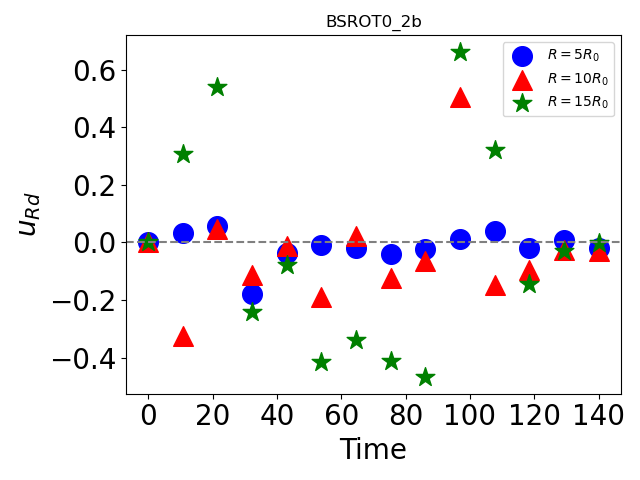}
\end{subfigure}

\begin{subfigure}{0.5\textwidth}
    \includegraphics[width=0.83\textwidth]{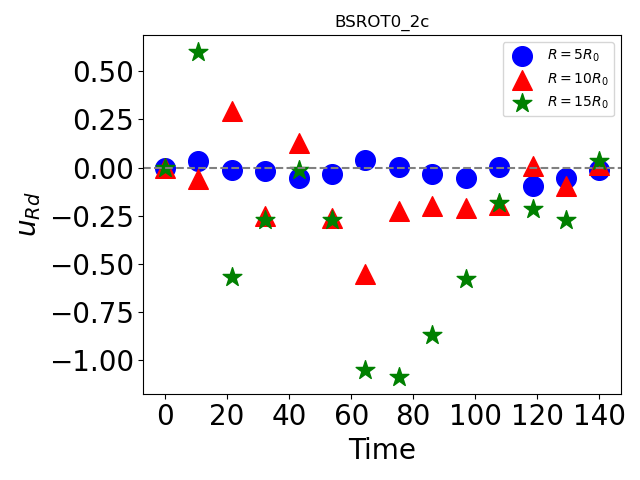}
\end{subfigure}
\begin{subfigure}{0.5\textwidth}
    \includegraphics[width=0.83\textwidth]{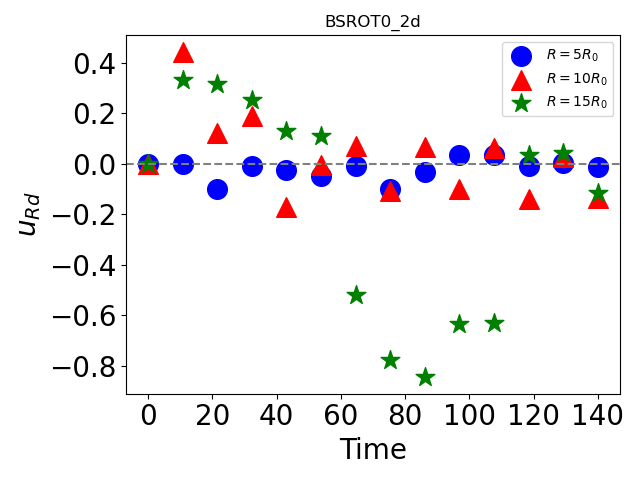}
\end{subfigure}
 
\caption{Radial velocity in the mid-plane as a function of time, calculated at three different points $R=5R_0$, $R=10R_0$ and $R=15R_0$. In all graphs presented has been considered that the rate of stellar rotation is $\delta_\star=0.2$. The corresponding model is indicated at the top of each subplot.}
\label{fig:radial_velocity}
\end{figure}

\begin{figure}
\begin{subfigure}{0.5\textwidth}
    \includegraphics[width=0.85\textwidth]{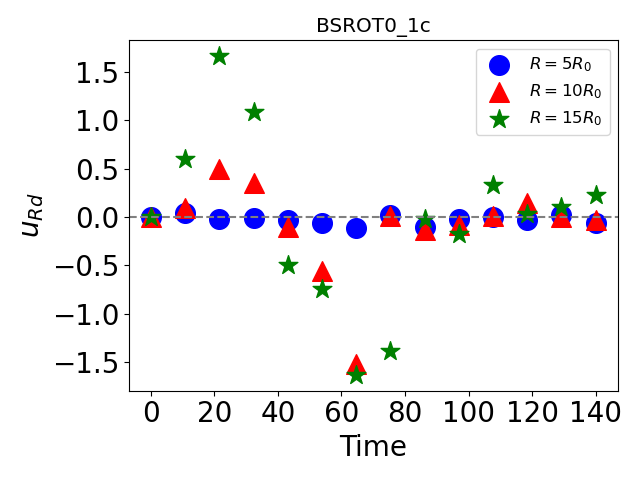}
\end{subfigure}
\begin{subfigure}{0.5\textwidth}
    \includegraphics[width=0.85\textwidth]{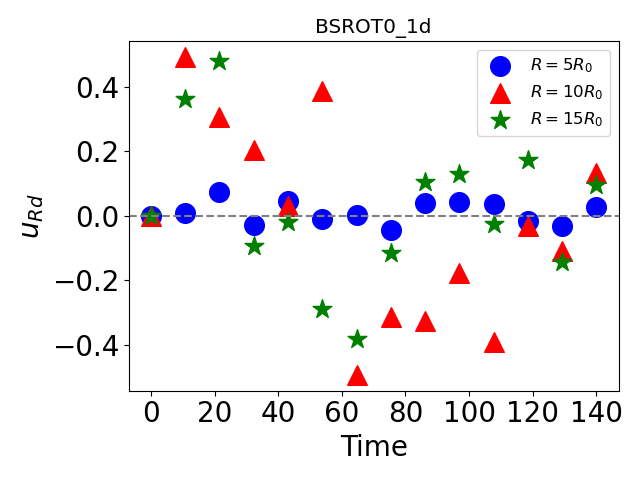}
\end{subfigure}
\caption{Radial velocity at the mid-plane as a function of time, calculated at three different points $R=5R_0$, $R=10R_0$ and $R=15R_0$. In all graphs presented has been considered that the rate of stellar rotation is $\delta_\star=0.1$. At the top of each subplot the model to which it corresponds is indicated.}
 \label{fig:vr01}
\end{figure}

\begin{figure}
\begin{subfigure}{0.5\textwidth}
    \includegraphics[width=0.95\textwidth]{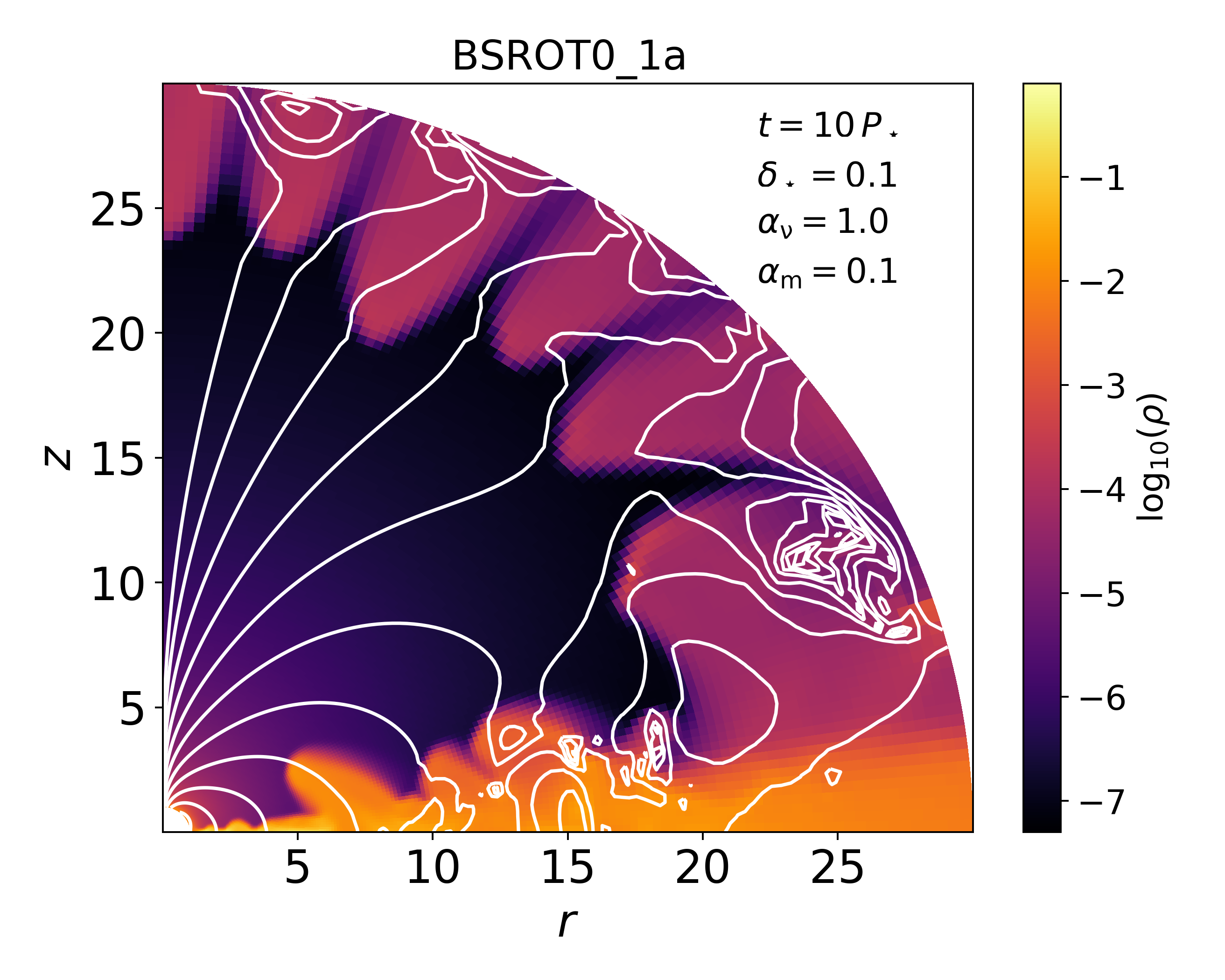}
\end{subfigure}
\begin{subfigure}{0.5\textwidth}
    \includegraphics[width=0.95\textwidth]{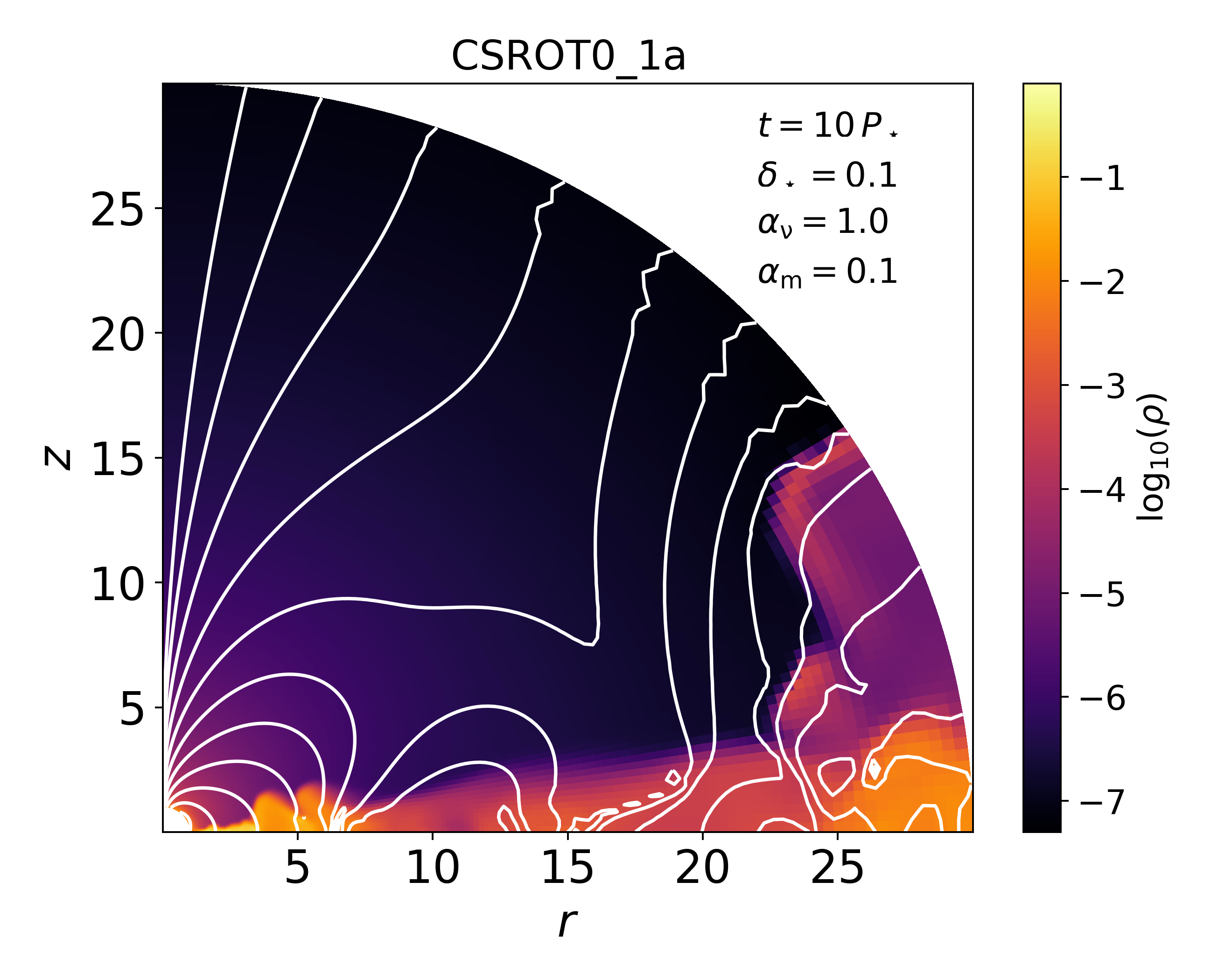}
\end{subfigure}
\begin{subfigure}{0.5\textwidth}
\includegraphics[width=0.95\textwidth]{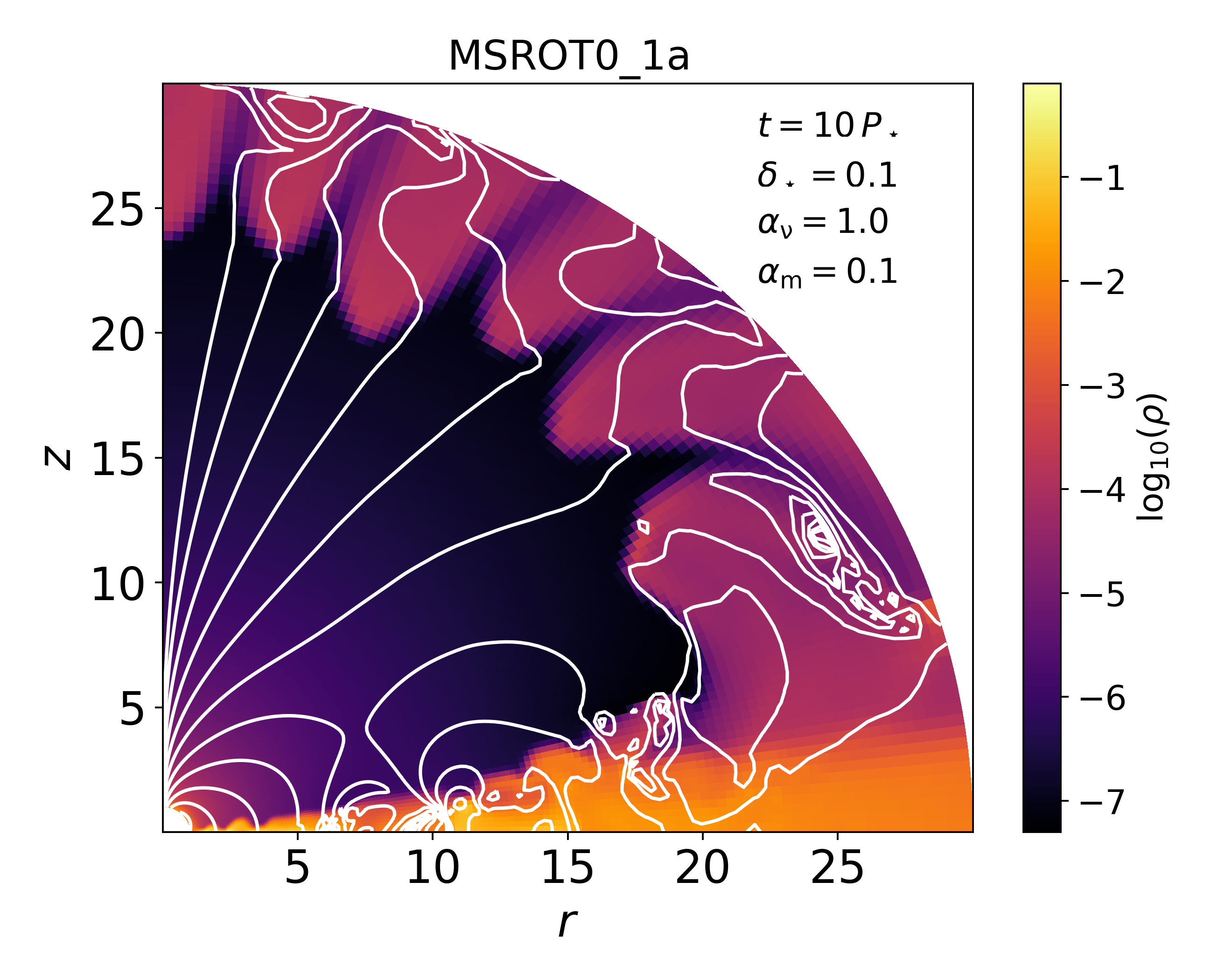}
    
\end{subfigure}
        
\caption{\textit{Comparison between different models.} In this figure we show the results using a time period of $10P_\star$, with a low stellar rotation rate $\delta_\star=0.1$, and a low magnetic diffusivity $\alpha_m=0.1$ using the three different descriptions to obtain the resistivity discussed in section \ref{sec:2.5}. The logarithmic gas density (color-scale background) changes from maximum value $\rho=-0.1$ in the disc to the minimum value $\rho=-7.3$ in the corona.The white lines show the poloidal magnetic field lines.}
\label{fig:models_comparison}
\end{figure}

\begin{table}
    \centering
    \begin{tabular}{p{1.5cm}| p{0.8cm}| p{0.8cm}| p{0.8cm}| p{0.8cm}|p{0.8cm}|}
        \hline
    Model & $\Omega_*/ \Omega_\mathrm{br}$ & $P_*$(days) & $\alpha_\mathrm{\nu}$& $\alpha_\mathrm{m}$ & $P_\mathrm{m}$ \\ 
    \hline

        CSROT0$\_$1a & $0.1$ & $4.6$ & $1.0$ & $0.1$ & $0.6$ \\
        CSROT0$\_$1b & $0.1$ & $4.6$ & $1.0$ & $1.0$ & $0.6$ \\
        MSROT0$\_$1a & $0.1$ & $4.6$ & $1.0$ & $0.1$ & $0.6$ \\
        MSROT0$\_$1b & $0.1$ & $4.6$ & $1.0$ & $1.0$ & $0.6$ \\
        BSROT0$\_$1a & $0.1$ & $4.6$ & $1.0$ & $0.1$ & $6.6$ \\
        BSROT0$\_$1b & $0.1$ & $4.6$ & $1.0$ & $1.0$ & $0.6$ \\
        BSROT0$\_$1c & $0.1$ & $4.6$ & $0.1$ & $0.1$ & $0.6$ \\
        BSROT0$\_$1d & $0.1$ & $4.6$ & $0.1$ & $1.0$ & $0.06$ \\
        BSROT0$\_$2a & $0.2$& $2.3$ & $1.0$ & $0.1$ & $6.6$ \\
        BSROT0$\_$2b & $0.2$& $2.3$ & $1.0$ & $1.0$ & $0.6$ \\
        BSROT0$\_$2c & $0.2$& $2.3$ & $0.1$ & $0.1$ & $0.6$ \\
        BSROT0$\_$2d & $0.2$& $2.3$ & $0.1$ & $1.0$ & $0.06$ \\

        \hline
    \end{tabular}
    \caption{Parameters of our numerical models. \textit{First column}. Name of each model, the first letter indicates what type of magnetic diffusivity was used (B: Bessolaz, C: Classic and M: Modified, respectively). \textit{Second column}. The stellar rotation rate. \textit{Third column}. Rotation period of the star. \textit{Fourth column}. The anomalous viscosity coefficient. \textit{Fifth column}. Dimensionless coefficient of magnetic diffusivity. \textit{Sixth column}. Prandtl number defined as $P_\mathrm{m}\equiv2\alpha_\nu/3\alpha_\mathrm{m}$.}
    \label{tab:Models}
\end{table}

\section{Discussion}
\label{sec:discussion}

\subsection{The mass flow}
The mass flow rate crossing the whole domain is the integral of the mass flux inside and outside of the disc. It is given by

\begin{equation}
    \Dot{M}= -4\pi R^2\int_0^{\pi/2} \rho u_{Rd} \sin{\theta} d\theta.
\end{equation}

Figure \ref{fig:massflow} shows the mass flow rate for four models (BSROT0\_2a, BSROT0\_2b, BSROT0\_2c and BSROT0\_2d), calculated at three different radius of the disc. The upper panel corresponds to $R=3.0R_0$. We find that for the four models the rate of mass flux towards the central star is always negative or zero with values between zero to -0.3. This behaviour was expected since in that part there is not formation of strong outflows (see figure \ref{fig:breaking} bottom panel). Also this means that there is no inflow of gas towards the central star in the mid-plane of the disc. The middle panel in Fig. \ref{fig:massflow} corresponds to the value given in equation (\ref{eq:Rt}) truncation radius $R=3.97R_0$. We can see some positive values in the models BSROT0\_2a, BSROT0\_2b and BSROT0\_2c. In the bottom panel in Fig. \ref{fig:massflow} we show the case when $R=10.0R_0$. Here we can see that for the models BSROT0\_2c and BSROT0\_2d the mass flux rate becomes positive after a time of $t=60T_0$ and $t=70T_0$, respectively, which means that the accretion disc stops at a radius greater than $R=10.0R_0$.

\subsection{Backflow and radial velocity}

Under some conditions, a phenomenon known as "backflow" occurs in the accretion disc \citep[see][and references therein]{MCK2023}. This phenomenon appears when a part of the disc flow moves in the opposite direction, away from the central object and usually appears in the mid-plane of the disc. In the figure \ref{fig:radial_velocity} we show the radial velocity for the models where the resistivity was obtained using the description given in Eq. (\ref{eq:Bess}) and considering the case when the rate of stellar rotation is $\delta_\star = 0.2$. 

In the plots we show the temporal evolution of the radial velocity component calculated at 
three different radii. The blue circles correspond to $R=5R_0$, the red triangles to $R= 10R_0$ and the green stars symbols correspond to $R= 15R_0$. In the four models we can observe that when the radial velocity is calculated in a radius corresponding to $R=5R_0$ the radial velocity is approximately zero over time. However, when the radius is taken as $R=10R_0$ or $R=15R_0$ we find negative values for the radial velocity component. That means an intermittent backflow appers as a function of the time, i.e., that backflow persist for certain time intervals and disapear in others.

On the other hand, for a stellar rotation rate of $\delta_\star=0.1$, considering a compact and extended magnetosphere (BSROT0$\_$1c and BSROT0$\_$1d models) we find again an intermittent backflow when we analyze the radial component of the velocity at radial distances from the star of $R=10R_0$ and $R=15R_0$, as can be seen in the figure \ref{fig:vr01}. It is important to point out that although in most of the BSROT0$\_$1 and BSROT0$\_$2 models the magnetic Prandtl number $P_\mathrm{m}$, is close to (or exceeds) the critical value of the Prandtl number  \citep[$P_\mathrm{m}^\mathrm{crit}\approx0.6$; see][]{MCK2023}, we find intermittent backflow at the mid-plane of the disc. Therefore, our results support the argument given in \citet{MCK2023} that the backflow is a physical effect rather than a numerical effect.

\subsection{Observed disc~structure in FS~CMas stars}

There are several observational studies in which the presence of thick disc structures around FS~CMa-type stars has been suggested. For example, the analysis of FS~CMa interferometric data by \cite{Hofmann22} shows that the outer dust ring has an inclination angle of approximately $40^{\circ}$. However, in the UV spectra a strong absorption of elements from the iron group is detected, indicating an enormous amount of gas at these angles.
Among other common properties of these objects, the presence of expanding layers, which can slow down, has also been suggested \citep{Kucerova13}.
Clumps moving away from the star were detected in \cite{Jerackova16} and evidence of material infall in \cite{Kucerova13}. 
Another important effect was reported by \cite{Ellerbroek15}. Based in the interferometric observations of HD~50138 in $Br\gamma$ with AMBER.They found that the star HD~50138 is surrounded by an au-scale rotating gas disc and spherical halo.  Among some of the results they reported are that the radius of the outer disc is $0.6\mathrm{au}$, the radius of the halo is $3\mathrm{au}$, and $\Delta v$ in the halo is $60 \mathrm{km/s}$. 
The results shown in our figure (\ref{fig:models_comparison}) can explain  observed phenomena reported by \citet{Ellerbroek15}. The models $\mathrm{BSROT0\_1a}$ and $\mathrm{MSROT0\_1a}$ show that the structure of the disc is thicker and also we can observe a structure on the edge that could be explained as a halo surrounding the disc.

Another study that supports the idea that FS~CMa-type stars are surrounded by thick disc was done by \cite{Zickgraf89}. They found that the FS~CMa stars are surrounded by a very extended and optically thick disc. Based on polarimetric observations at Calar-Alto, they found that the disc opening angle is $\approx 30^{\circ}$ for the stars MWC~645 and MWC~939. We find that in all our models there is an increase in the aspect ratio as it evolves over time. For instance, in the case of the $\mathrm{BSROT0\_1b}$ model the aspect ratio increased 3.5 times for a corresponding time of $P_\star=130$ stellar periods that means that in this case we found that the disc opening angle is $\approx 25^{\circ}$, while for the $\mathrm{BSROT0\_2b}$ model the disc opening angle is $\approx 33^\circ$  i.e. the aspect ratio increased 4.6 times for the same time period. In Fig. \ref{fig:breaking} one can see how the disc is thickened.

\subsection{Magnetic field configuration inside the accretion disc}

In the case of a compact magnetosphere (models with $\alpha_\mathrm{m}=0.1$), we find that the magnetic field configuration is far from the standard configuration of a vertical magnetic field (resulting from an initial dipole configuration). We find that the poloidal magnetic field lines are twisted (see for example Fig.\ref{fig:diff}). In such configuration, the magnetic reconection heating the corona is highly probable, that could explain the presence of the weak Raman lines found observationally in
several FS~CMa stars. Additionally, we can see in the upper panel in Fig. \ref{fig:diff} that the magnetic field lines within the disc inflate and twist rapidly (within the first 10 orbital periods of the star), which gives rise to the formation of knots at different radial distances near the disc mid-plane. These knots substantially modify the density of the disc producing regions of low density. In other words, they can generate the formation of shallow gaps in the disc which can be displaced radially depending on the inflation rate of the magnetic field lines (see upper panel in Fig. \ref{fig:models_comparison}). In addition, the knots can also move towards the corona region, dragging gas with them from the surface of the disc, so the disc suffers a thickening. It should be emphasized that regardless of the analytical form of the magnetic diffusivity described by equations (\ref{eq:nuC}-\ref{eq:Bess}), when $\alpha_\mathrm{m}=0.1$ we find the formation of at least a pair of knots tied to the mid-plane of the disc. This result implies the generation of strong turbulence within the disc which can modify the velocity components and the density distribution.

On the other hand, when $\alpha_\mathrm{m}=1$ we find that the formation of knots occurs in the corona region near the disc layer (see lower panel in Fig. \ref{fig:diff}), and therefore the perturbations in density and velocities within the disc are smaller. However, these perturbations produce a higher accretion rate in the mid-plane of the disc (see below) unlike other studies where strong magnetic fields are also considered \citep{MKS1997,Romanova_etal2002,CemKP2023}. We think that the main reason for obtaining a greater perturbation in density within the disc is due to the vertical dependence of the density. Since we use a vertical profile of the density that depends on the height through the polar $\theta$-coordinate and, therefore the density of the disc decays for values of $\theta$ close to the aspect ratio of the disc $h$ and facilities the inflation of the magnetic lines. Otherwise, as shown in \citet{CemKP2023} when a density-taper function ($\rho=\rho_c r^{-3/2}$) in the corona-disc interface region is used, the density becomes more noisy there without drastic density changes near the mid-plane since the magnetic field decreases by at least two orders of magnitude.

\section{Summary and Conclusions}
\label{sec:conclusions}

We have performed a set of $2.5D$  magnetohydrodynamical simulations to analyze the star-disc interaction of FS~CMa type stellar objects that exhibit a strong magnetic field at the star surface \citep[$B_\star=6.2kG$;][]{Korcakova22}. Based on the results of Paper I we set the value of the disc density $\rho_{d0}=10^{-13}\mathrm{gcm}^{-3}$, and we have considered as free parameters; the viscosity $\alpha_\nu$, and the resistivity  $\alpha_\mathrm{m}$ inside of the accretion disc, as well as the stellar rotation rate $\delta_\star$. By varying the resistivity, we analyze the effect of the magnetic field on the density of the disc and the corona region in the cases of a compact ($\alpha_\mathrm{m}=0.1$) and extended magnetosphere ($\alpha_\mathrm{m}=1$), respectively.

We find that a higher stellar rotation rate ($\delta_\star=0.2$) can modify the density structure in the internal region of the disc, generating different layers when $R\leq10R_0$, which adds to the effect generated by a higher viscosity ($\alpha_\nu=1$) that produces the thickening of the accretion disc. In the case of a low-viscosity ($\alpha_\nu=0.1$), we find that in both models star-disc interaction with a compact or extended magnetosphere the fall of gas towards the star can be in the mid-plane. However, we do not rule out that the funnel effect may occur where the accretion of gas is conducted near the star's poles, since until the end of our simulations, the density structure in most of our models resembles a form similar to the funnel effect reported in less massive stars with a lower magnetic field \citep[see for instance][]{Romanova_etal2002,Bessolaz07,Zanni2009,ZF2013,Cemelij19}. 

On the other hand, since we consider a value of $\alpha_\nu=0.1<0.685$, we have analyzed if there is a backflow in the mid-plane of the disc which was previously reported \citep[see][and references therein]{Cemelij19}. We find that regardless of the type of magnetosphere modeled there is intermittent backflow, as well as several knots forming in the magnetic field lines near and in the mid-plane of the disc. These knots produce perturbations in the density and velocity components, as well as the formation of shallow gaps in the density whose position depends on the inflation of the magnetic field lines.

Furthermore, we conclude that our findings may have
direct implications for understanding the observational results for the study of post-mergers among FS~CMa stars, because as we can see in our simulations the disc structure changes and becomes thicker significantly.
Our simulations could also explain some other properties that have been found observationally in FS~CMA stars \citep{Kucerova13}, such as: material outflow and inflow, the formation of layers, and also the presence of the weak Raman lines found in several FS~CMA stars than can be explained by corona heating due to the magnetic reconection.

\section*{Acknowledgements}

We thank the referee for his/her thorough reading of
our manuscript, and for his/her constructive and careful report, which helped us improve the previous version of this work. This work was supported by the Czech Science Foundation (grant 21-11058S). The work of ROC was supported by the Czech Science Foundation (grant 21-23067M). Computational resources
were available thanks to the Ministry of Education, Youth and Sports of the Czech Republic through the e-INFRA CZ (ID:90254).

\section*{Data Availability}

 The PLUTO code is available from \href{http://plutocode.ph.unito.it}{http://plutocode.ph.unito.it}. The input files for generating our 3D magnetohydrodynamical simulations will be shared on reasonable request to the corresponding author.



\bibliographystyle{mnras}
\bibliography{example} 




\appendix


\section{Funnel effect in low-magnetized stars: Set-up comprobation}
\label{sec:appendixB}
\subsection{Test code for low-magnetized stars}

\begin{figure}
\includegraphics[width=\columnwidth]{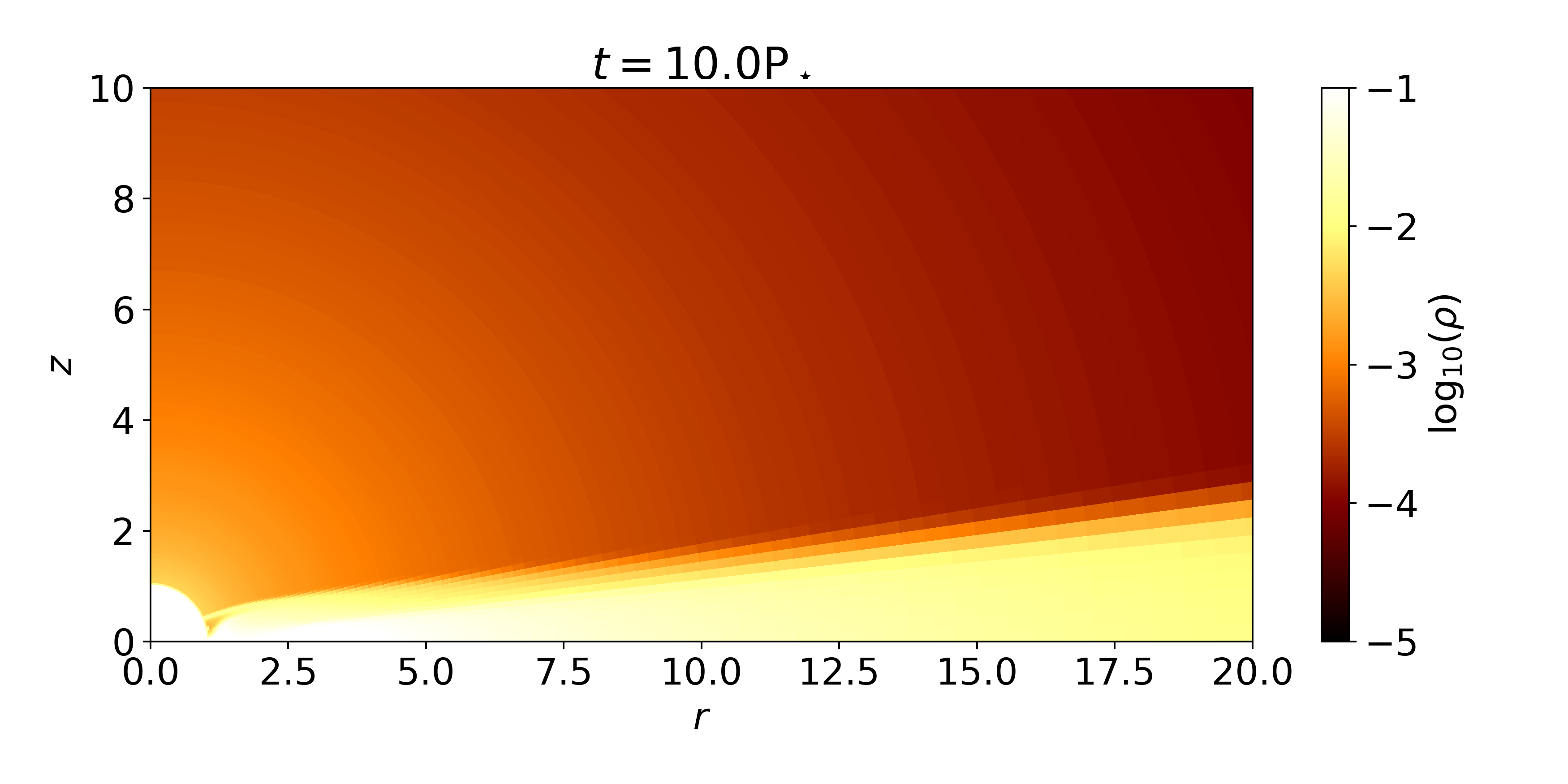}
    \caption{ Density in the hydrodynamic model  in logarithmic scale in units of $\rho_{0}$}
    \label{fig:apendix1}
\end{figure}

\begin{figure}
\includegraphics[width=\columnwidth]{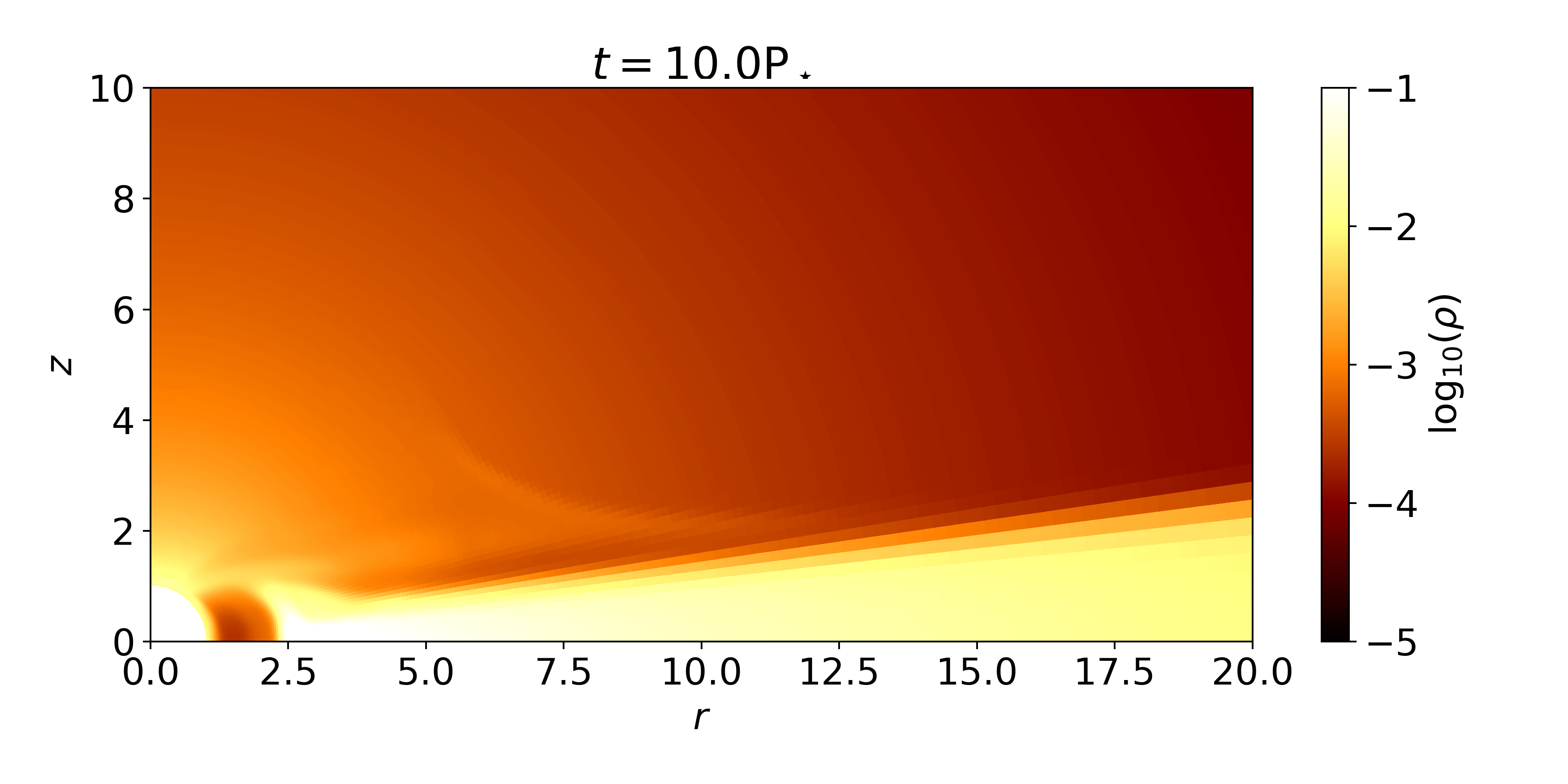}
    \caption{ Density in the magnetohydrodynamical model  in logarithmic scale in units of $\rho_{0}$}
    \label{fig:apendix2}
\end{figure}

In Figure \ref{fig:apendix1} we show the results of gas density at $t=10P_\star$ obtained from the simulations considering the hydrodynamic case (i.e. without magnetic field) and a rotation of the star of 0.1. It is clearly observed that the material is being accreted towards the center. On the other hand in fig. \ref{fig:apendix2} we show the results obtained from the simulations considering the magnetohydrodynamic case for a star with a rotation of 0.1, in this case we used a weak magnetic field ($B_\star =140G$) in order to see if our results are in agreement  with others studies where they report the ocurrence of the funnel effect in T-Tauri stars. In fig. \ref{fig:apendix2} we shown the gas density distribution of our numerical test where we can see  that from early times ($t=10P_\star$) the funel effect is already formed. This means, that the matter flows out of the disc plane and essentially free-falls along the stellar magnetic field lines. Therefore, our results are in agreement with what was previously reported by \cite{Zanni2009}, \cite{Romanova_etal2002}, \cite{Cemelij19}. 

\bsp	
\label{lastpage}
\end{document}